\documentclass[twocolumn]{revtex4}
\usepackage{graphicx}

\begin{document}

\preprint{APS}

\title{Soft x-ray resonant diffraction study of magnetic and orbital correlations in a manganite near half-doping}

\author{K. J. Thomas$^{1}$}
\author{J. P. Hill$^{1}$}
\author{Y-J. Kim$^{1}$}
\author{S. Grenier$^{1,2}$}
\author{P. Abbamonte$^{3}$}
\author{L. Venema$^{4}$}
\author{A. Rusydi$^{3}$}
\author{Y. Tomioka$^{5}$}
\author{Y. Tokura$^{6}$}
\author{D. F. McMorrow$^{7}$}
\author{M. van Veenendaal$^{8}$}

\affiliation{
$^{1}$ Department of Physics, Brookhaven National Laboratory,  Upton, New York  11973 \\ 
$^{2}$ Department of Physics and Astronomy, Rutgers University, Piscataway, New Jersey  08854 \\
$^{3}$ National Synchrotron Light Source, Brookhaven National Laboratory, Upton, New York  11973\\ 
$^{4}$ Materials Science Centre, University of Groningen, 9747 AG Groningen, The Netherlands \\ 
$^{5}$ Correlated Electron Research Center (CERC), National Institute of Advanced Industrial Science and Technology (AIST), Tsukuba 305-8562, Japan\\ 
$^{6}$ Department of Applied Physics, University of Tokyo, Tokyo 113-8656, Japan and Spin Superstructure Project, ERATO, Japan Science and Technology Corporation (JST), Tsukuba 305-8562, Japan \\ 
$^{7}$ Materials Research Department, Ris\o \ National Laboratory, 4000 Roskilde, Denmark\\ 
$^{8}$ Northern Illinois University, Dekalb, Illinois 60115 and Argonne National Laboratory, 9700 South Cass Avenue, Argonne, Illinois 60439}

\date{\today}

\begin{abstract}
We have utilized the resonant x-ray diffraction technique at the Mn $L$-edge in order to directly compare magnetic and orbital correlations in the Mn sub-lattice of Pr$_{0.6}$Ca$_{0.4}$MnO$_{3}$.  The resonant line shape is measured below $T_{OO}$ $\sim$ 240 K at the orbital ordering wave vector (0,$\frac{1}{2}$,0), and below T$_{N}$ $\sim$ 175 K at the antiferromagnetic wave vector ($\frac{1}{2}$,0,0).  Comparing the width of the super-lattice peaks at the two wavevectors, we find that the correlation length of the magnetism exceeds that of the orbital order by nearly a factor of two.  Furthermore, we observe a large ($\sim$ 3 eV) shift in spectral weight between the magnetic and orbital line shapes, which cannot be explained within the classic Goodenough picture of a charge-ordered ground state.   To explain the large shift, we calculate the resonant line shapes for orbital and magnetic diffraction based on a relaxed charge-ordered model. 
\end{abstract}

\pacs{Valid PACS appears here}

\maketitle

\newcommand*{\mnthree}{Mn$^{3+}$}
\newcommand*{\mnfour}{Mn$^{4+}$}
\newcommand*{\pcmo}{Pr$_{1-x}$Ca$_{x}$MnO$_{3}$}
\newcommand*{\pcmofour}{Pr$_{0.6}$Ca$_{0.4}$MnO$_{3}$}
\newcommand*{\too}{$T_{OO}$}
\newcommand*{\tn}{$T_{N}$}
\newcommand*{\tco}{$T_{CO}$}

\begin{figure}
\includegraphics[height=0.22\textheight,width=0.5\textwidth]{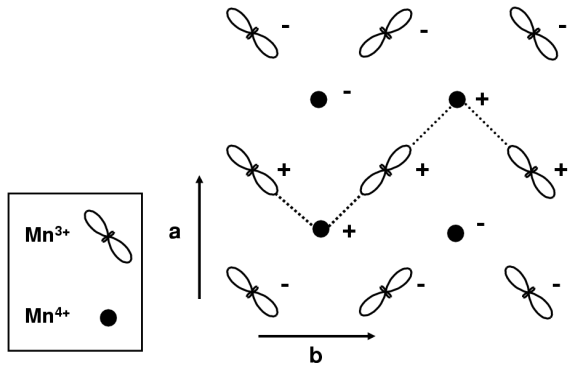}
\caption{CE magnetic ground state for half-doped manganites.  The plus and minus signs denote spin directions.  In \pcmofour, magnetic scattering occurs along \textbf{a} at ($\frac{1}{2}$,0,0) reflections and orbital scattering occurs along \textbf{b} at (0,$\frac{1}{2}$,0), reflections.}
\label{fig:figure1}
\end{figure}

In a number of manganites, including Pr$_{1-x}$Ca$_{x}$MnO$_{3}$, La$_{1-x}$Ca$_{x}$MnO$_{3}$ and La$_{2-x}$Sr$_{x}$MnO$_{4}$, the dynamics resulting in a charge-ordered, insulating state in the vicinity of half-doping are still not well understood.  In his seminal work on exchange interactions in manganites, Goodenough considered charge and orbital order at half-doping as a precursor to the magnetic CE ground state\cite{goodenough}.  In this picture, charge ordering at $T_{CO}$ results in a checkerboard pattern of \mnthree \ and \mnfour \ sites (Figure 1).  The \mnthree \ sites each have one $e_{g}$ electron and are thus Jahn-Teller (JT) active, while the $e_{g}$ levels are empty on the \mnfour \ sites.  A cooperative JT distortion and orbital ordering of the \mnthree \ sites at \too \ = \tco \ occurs concomitantly with the charge ordering.  The in-plane JT distortions favor occupation of $3x^{2}$ - $r^{2}$ and 3$y^{2}$ - $r^{2}$ orbitals on the \mnthree \ sites, establishing a ferromagnetic exchange pathway along orbital zig-zag chains (dotted line in Figure 1).  Magnetic CE ordering, defined by antiferromagnetically coupled ferromagnetic chains, occurs at \tn \ $\leq$ \too.    

Extensive neutron and x-ray scattering measurements support the Goodenough picture\cite{radaelli_prb, jirak, moritomo_lsmo4, sternlieb, zimmermann_prb}.  However, this model makes definite predictions about the Mn ground state for which experimental confirmation is still lacking.  One controversial issue is the degree to which the ``\mnthree'' and  ``\mnfour'' sites are actually separated by unit valence.  Indeed, a number of experimental results challenge the charge ordering model\cite{garcia,zener_polaron}.  Recent analysis of Mn $K$-edge resonant x-ray diffraction in \pcmofour \ suggests that the charge ordering is far from complete, but supports the presence of an orbitally ordered ground state\cite{grenier_condmatt}.   In addition, the Goodenough model predicts a strong coupling between the orbital and magnetic correlations within the \mnthree \ sub-lattice.  However, no single experimental technique has been able to directly compare magnetic and orbital correlations.  

Recently, it has been demonstrated that resonant diffraction at the Mn $L$-edges may be utilized to study magnetic and orbital order in the manganites\cite{wilkins1,wilkins2,castleton}.  The Mn $L$-edge resonances involve strong dipole transitions from the \textit{2p} core levels to unoccupied states within the \textit{3d} band.  The resonant enhancement of magnetic scattering at the Mn $L$-edge is significantly greater than at the $K$-edge\cite{hannon_trammel}.  Furthermore, the structure of the resonant diffraction line shape provides a direct spectroscopy of the Mn $3d$ band in the presence of correlations.   

In this letter, we present a resonant x-ray diffraction study of a near half-doped manganite at the Mn $L$-edge which permits the first direct comparison between magnetic and orbital correlations.  Our data provide evidence against a charge ordering into distinct \mnthree \ and \mnfour \ sub-lattices in \pcmofour, and indicate a more complex ground state.  First, we find that the magnetic correlations are significantly longer-ranged than the orbital correlations.  Second, the spectral line shapes and the difference in intensity between magnetic and orbital scattering disagree with the ionic CE picture.  Rather, we suggest that while the $e_{g}$ electron is still localized, it is shared between neighboring lattice sites\cite{MvV}. 

We have focused on twinned, single crystal \pcmofour, which has been described as a CE-type antiferromagnet\cite{jirak}. \pcmofour \  exhibits charge/orbital order at $T_{OO}$ = $T_{CO}$ $\sim$ 240 K and long-range spin order at \tn \ $\sim$ 170 K, which result in super-lattice reflections at $(0,\frac{1}{2},0)$ and $(\frac{1}{2},0,0)$, respectively, with respect to the orthorhombic unit cell ($a$ = 5.41 \AA \ and $b$ = 5.43 \AA\ ).  The twinned crystal has both [100] and [010] oriented domains at the sample surface such that both  ($\frac{1}{2}$,0,0) and (0,$\frac{1}{2}$,0) reflections are accessible at the same scattering angle. At the Mn $L$-edges, which are $\sim$ 650 eV and lie in the soft x-ray regime,  $2\theta$ $\sim$ 124$^{\circ}$ for these reflections.  Although the ($\frac{1}{2}$,0,0) and (0,$\frac{1}{2}$,0) reflections are not resolved in momentum space, the magnetic and orbital scattering are clearly distinguished by different transition temperatures and correlation lengths.

\begin{figure}
\includegraphics[height=0.4\textheight,width=0.4\textwidth]{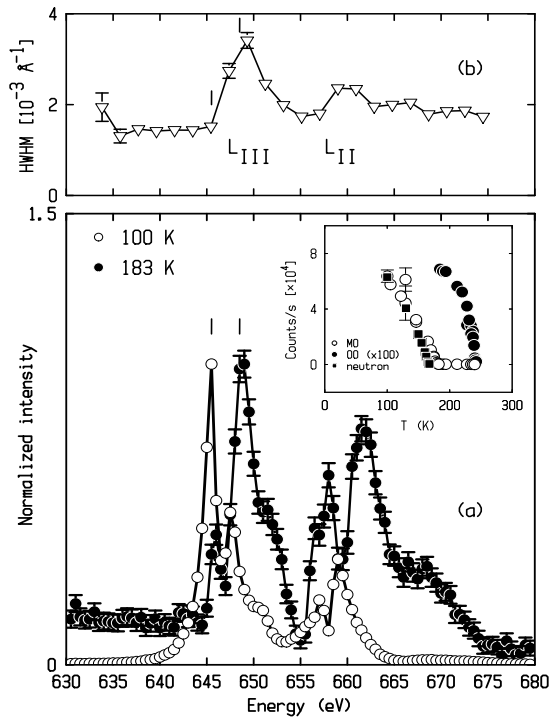}
\caption{(a) Energy scans at fixed momentum transfer at $T$ = 100 K ($T$ $<$ $T_{N}$)  and $T =$ 183 K ($T_{N}$ $<$ $T$ $<$ $T_{OO}$). (Inset) Temperature dependence of the peaks in the magnetic (MO, open circles) and orbital (OO, closed circles) spectra and the ($\frac{3}{2}$, 0, 0) magnetic reflection measured with neutrons (dark squares). (b) Energy dependence of the \textit{HWHM} of the ($\frac{1}{2}$,0,0) magnetic peak. The tick marks in (a) and (b) denote the peaks in the magnetic and orbital spectra.}  
\label{fig:comp_energyspectra}
\end{figure}

The diffraction measurements were performed at the X1B undulator beamline at the National Synchrotron Light Source.  The spectrometer is housed in an UHV chamber to reduce adsorption on the sample surface.  A focusing gold diffraction grating selects the incident photon energy.  Variable width  slits before and after the grating determine the energy resolution, which is $<$ 0.5 eV (calculated) for the measurements shown here. The  incident beam is $\pi$-polarized and diffraction occurs in the horizontal geometry. The sample is mounted to the cold finger of a He cryostat. The azimuth is fixed with the \textit{c}-axis normal to the scattering plane.

Figure 2(a) shows the energy dependence of the diffracted intensity at \textbf{Q} = $(\frac{1}{2},0,0)$/$(0,\frac{1}{2}, 0)$ in the vicinity of the Mn $L_{III}$ ($2p_{\frac{3}{2}} \rightarrow 3d$) and $L_{II}$ ($2p_{\frac{1}{2}} \rightarrow 3d$) absorption edges.  The scans were taken at 183 K,  below $T_{OO}$ but above \tn \ , and at 100 K, below \tn.  A high temperature ($T$ = 240 K) spectrum has been subtracted from both spectra to remove the contribution from specular reflectivity and fluorescence.  To emphasize the difference in the line shapes, the two curves are scaled to the same intensity at their respective maxima.  As we discuss below, the peak in the 100 K spectrum ($\sim$ 60,000 counts/s) is actually 100 times more intense than that of the 183 K spectrum.  The background subtraction is thus only significant for the orbital spectrum, for which it is approximately a 10\% effect.

There is a dramatic shift in the resonant spectral weight upon warming through \tn: Below \tn \ there is a sharp peak at 645.5 eV within the $L_{III}$ region and smaller scattering within the $L_{II}$.  Above \tn, the peak in the $L_{III}$ occurs at 648.5 eV and the $L_{III}$ and $L_{II}$ scattering are nearly equal in intensity.  Warming from 100 K to \tn \ results in no change in the line shape; only a decrease in intensity is observed.  Similarly, above \tn \ the line shape remains constant, exhibiting an order-parameter like drop-off in intensity around 240 K.  The inset to Figure 2(a) shows the temperature dependencies of the intensity measured at 645.5 eV and 648.5 eV. (The intensity of the 648.5 eV feature is only indicated above 180 K, as the contribution from the magnetic scattering dominates at lower temperatures.)  The transition temperatures $T_{N}$ = 175 K and $T_{OO}$ = 240 K agree well with those measured in Ref. \cite{jirak} and comparison with neutron data on the same sample\cite{youngjune_unpublished} confirms that the resonant diffraction intensity below $T_{N}$ is indeed sensitive to the magnetic ordering.  

In Figure 2(b), we show the \textit{HWHM} of the magnetic Bragg peak as a function of energy near the Mn $L$-edges.  Away from the $L_{II}$ and $L_{III}$ edges, photoelectric absorption limits the penetration depth to $\sim$ 1000 \AA.  Figure 2(b) indicates the further decrease in penetration depth as the incident energy is tuned through the Mn $L$-edges, resulting in a broadening of the magnetic peak.  Comparing the absorption and the resonant diffraction spectra shows that the peak in the absorption coincides with the peak in the orbital spectrum.  We note that the resonant line shapes in 2(a) are not corrected for absorption effects; however, calculations of the absorption away from the $L$-edge suggest that the change in absorption does not significantly affect the resonant line shapes.

\begin{figure}
\includegraphics[height=0.2\textheight,width=0.45\textwidth]{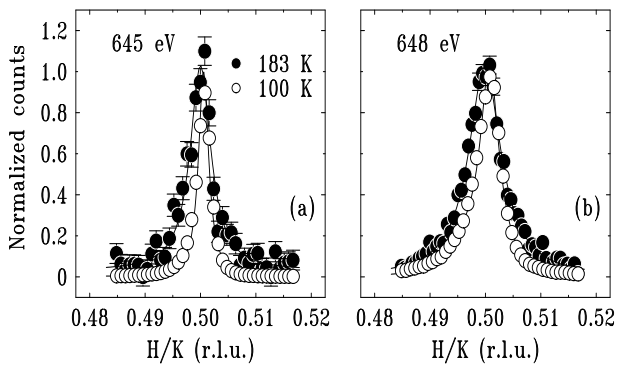}
\caption{Longitudinal momentum scans above ($T$ = 183 K, dark circles) and below ($T$ = 100 K, open circles) T$_{N}$ at (a) 645 eV and (b) 648 eV}
\label{fig:tdependence}
\end{figure}

This study permits the first direct comparison of magnetic and orbital correlations in a manganite.  Figure 3 compares longitudinal momentum scans through the orbital and magnetic diffraction peaks at 645 and 648 eV.  At \textit{both} energies, the orbital peak is significantly broader than the magnetic peak, indicating that the difference in width results from differences in the magnetic and orbital correlation lengths and not absorption or extinction effects. We define the inverse correlation length, $\frac{1}{\xi} =$ \textit{HWHM}, where the \textit{HWHM} of the peak is fit with a Lorentzian line shape.  Analyzing the peaks at 645 eV, we find an orbital correlation length $\xi_{b}^{orb}$ = 375 $\pm$ 40 \AA.  The magnetic correlation length however, is more than a factor of two longer with $\xi_{a}^{mag}$ = 780 $\pm$ 40 \AA.

In earlier $K$-edge resonance experiments on \pcmo, it was found that the orbital correlation length was $\sim$ 300 \AA \ for $x$ = 0.4, in good agreement with our results, and $\sim$ 160 \AA \ for $x$ = 0.5\cite{hillappphys}.  Comparison with neutron scattering measurements, which showed similarly short-ranged magnetic correlations in the \mnthree  \ sub-lattice, led to suggestions that the \mnthree \ magnetic and orbital correlation lengths were identical \cite{radaelli_prb, jirak_prb2000}. These results supported the CE picture, since orbital domain walls create magnetic domain walls in the \mnthree \ sub-lattice.  However, the $L$-edge resonant diffraction at ($\frac{1}{2}, 0, 0)$ is only sensitive to the \mnthree \ sites and clearly indicates longer range correlations between spins than orbitals within this sub-lattice.  Furthermore, we confirmed that the discrepancy between magnetic and orbital correlation lengths is isotropic in the $ab$ plane by measuring transverse scans through the magnetic and orbital peaks.  This measured difference between orbital and magnetic correlation lengths presents a direct challenge to the Goodenough picture of magnetic and orbital coupling.

A further challenge to the Goodenough picture lies in the comparison between the magnetic and orbital resonant diffraction line shapes and intensities.  In the CE model, the magnetic and orbital scattering at ($\frac{1}{2}$,0,0) and (0,$\frac{1}{2}$,0) result entirely from the \mnthree \ sub-lattice.  However, a single atom picture cannot explain two of our key findings.  

First, the magnetic scattering is significantly stronger than the orbital scattering.  A larger population of \textit{a}-oriented than \textit{b}-oriented domains could favor magnetic scattering.  In a separate experiment, we found that the orbital and magnetic spectra were maximized for different positions of the beam on the sample surface.  By optimizing the beam position separately for the orbital and magnetic spectra, we found a ratio of magnetic to orbital peak intensities $\sim$ 25:1.  Other factors will also contribute to a smaller orbital scattering, such as the shorter in-plane correlation length and the greater absorption at the orbital peak energy.    

Second, the orbital resonance line shape is peaked at energies where there is little spectral weight in the magnetic scattering. This is difficult to reconcile with a single atom picture, where the spectral weight for magnetic and orbital scattering is expected to be in the same energy range.

However,  we find we can describe both the resonant line shapes and difference in intensities by assuming a more realistic ground state in which the $e_{g}$ electron is shared between neighboring sites,
\begin{eqnarray}
|g\rangle = \alpha |3;4\rangle + \beta |4;3\rangle
\label{eqn:one}
\end{eqnarray}
Here, $|i;j\rangle$ denotes $i$ electrons on the JT distorted site where the scattering takes place and $j$ electrons on a neighboring site along a ferromagnetic chain.  (In the CE model depicted in Figure 1, $|g\rangle = |4;3\rangle$.)   The hopping parameter $t$ that couples the two configurations in Eqn. \ref{eqn:one} is approximately of the order of 1-1.5 eV and is therefore significantly larger than the JT energy of 0.1-0.3 eV.  We can therefore expect $\alpha \cong \beta$.  

The inclusion of coupling to a neighboring site leads to two different (dipole allowed) intermediate states:
\begin{eqnarray}
|n_{1}\rangle & = & \beta^{\prime}|\underline{c}5;3\rangle - \alpha^{\prime}|\underline{c}4;4\rangle \\
|n_{2}\rangle & = & \alpha^{\prime}|\underline{c}5;3\rangle + \beta^{\prime}|\underline{c}4;4\rangle
\end{eqnarray}
where ${\underline c}$ denotes a $2p$ core hole.  Within an ionic picture, the energy difference $\Delta_1=E_{|\underline{c}5;3\rangle} - E_{|\underline{c}4;4\rangle}$ is related to the Coulomb interactions between $3d$ electrons, $U_{dd}$, and between the $3d$ electrons and the $2p$ core hole, $U_{pd}$.  Thus,  $ \Delta_{1} = U_{dd} - U_{pd}$, which can be of the order of a few eV. 

We will argue that the magnetic scattering is dominated by the $|3;4\rangle$ configuration of the ground state while the orbital scattering is dominated by the $|4;3\rangle$ configuration, and that the orbital scattering is dominated by transitions to the higher energy $|n_{2}\rangle$ intermediate state, while the magnetic scattering is dominated by transitions to the lower energy $|n_{1}\rangle$ intermediate state.  Thus, the orbital scattering spectral weight is shifted to a higher energy than the magnetic scattering.

\begin{figure}
\includegraphics[height=0.4\textheight,width=0.4\textwidth]{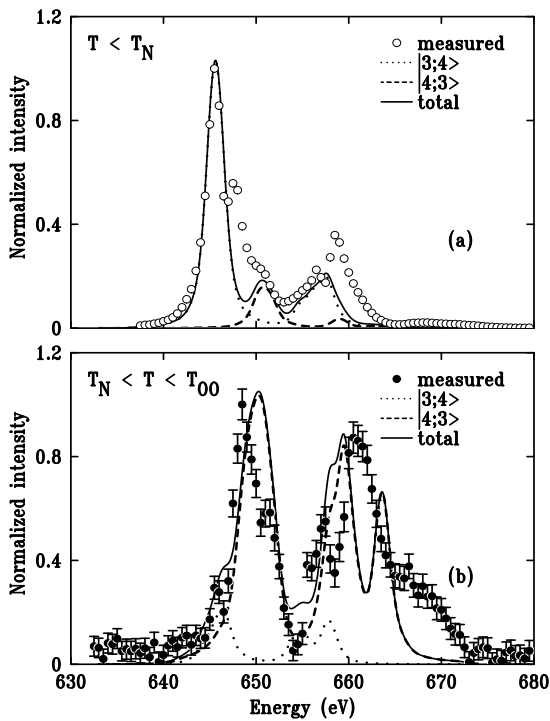}
\caption{Comparison between the calculated and measured resonance line shapes. In (a) the magnetic scattering is dominated by the $|3;4\rangle$ configuration while in (b) the orbital scattering is dominated by the $|4;3\rangle$ configuration.}
\label{fig:calclineshapes}
\end{figure}

The ground-state configurations $|3;4\rangle$ and $|4;3\rangle$ make significantly different contributions to the orbital scattering.  Contributions to orbital scattering occur in two ways.  First, the JT distortions split the $3d$ intermediate states, leading to a finite orbital scattering on resonance.  This effect occurs for both intermediate state configurations $|\underline{c}4;4\rangle$ and $|\underline{c}5;3\rangle$.  However, as shown in Ref. \cite{castleton}, this effect is only important if the JT splitting is significantly larger than the core-hole lifetime broadening of 0.3-0.5 eV.  A second and much stronger effect occurs if the lower $e_{g}$ level on the JT site is occupied in the ground state.  Therefore, the $|4;3\rangle$ configuration makes the dominant contribution to the orbital scattering.

In contrast, there is no reason to assume that the magnetic scattering of the $|3;4\rangle$ and $|4;3\rangle$ configurations differ strongly in intensity (although their spectral line shapes can be different).  However, it is straightforward to show that the ratio of scattering intensities for the two different intermediate states $|n_{1}\rangle$ and $|n_{2}\rangle$ is $r=[(\alpha \alpha^{\prime} + \beta \beta^{\prime})/(\alpha \alpha^{\prime} - \beta \beta^{\prime})]^{4}$.  Assuming $\alpha \cong \beta$, $t$ = 1.1 eV and a 3.5 eV splitting between $|n_{1}\rangle$ and $|n_{2}\rangle$, we obtain $r=20$.  This shows that the scattering through intermediate states $|n_{2}\rangle$, and therefore the orbital scattering, is about 20 times weaker than the magnetic scattering.  Similarly, the magnetic scattering arising from $|n_{2}\rangle$ will also be significantly smaller than that from $|n_{1}\rangle$.  

To quantify these arguments, we calculated the magnetic and orbital scattering in the ordered state by exact diagonalization of a Hamiltonian including the full $U_{pd}$ and $U_{dd}$ Coulomb interaction and the $2p$ and $3d$ spin-orbit coupling for the different configurations.  We assumed an octahedral crystal field of 1.5 eV and a JT splitting of the $e_{g}$ orbitals of 0.1 eV.  

The results are compared with the measured spectra in Figure 4, which shows the separate and combined contributions from the $|3;4\rangle$ and $|4;3\rangle$ configurations.  Figure 4(a) shows the magnetic scattering, which is dominated by the $|3;4\rangle$ configuration.  The $|4;3\rangle$ contribution is strongly reduced by the spectral weight effect described above.  Figure 4(b) shows the orbital spectrum.  It is dominated by the $|4;3\rangle$ configuration, which though reduced by the same spectral weight transfer compared to the magnetic scattering, still contributes more strongly than the $|3;4\rangle$ configuration to the orbital scattering.  

We note that this model is preliminary and that a number of open questions remain.  For example, the model cannot explain why the magnetic spectrum is strongest below the peak in the absorption.  Nevertheless, we feel it is a step in the right direction and illustrates the quality of ground state information obtained with this technique.

In summary, we have measured the resonant line shape at the Mn $L$-edge in \pcmofour, at \tn \ $<$ $T$ $<$ \too \ and at $T$ $<$ \tn.  The comparison of the magnetic and orbital line shapes, in addition to theoretical calculations, strongly suggest that the Mn sites are highly mixed between the \mnthree \ and \mnfour \ formal oxidation states.  Furthermore, we find that the magnetic correlations are longer ranged than the orbital correlations.  These findings pose a direct challenge to the classic Goodenough picture of charge, orbital and magnetic order in half-doped manganites.  The results presented here emphasize the  sensitivity of resonant diffraction at the Mn $L$-edge to the ground state of the $3d$ electrons and encourage a new approach to theoretical calculations.  

We thank J. van den Brink for comments.  The measurements performed at BNL were supported by US DOE, Division of Materials Science under contract no. DE-AC02-98CH10886.  Work by M. v. V was supported by DOE contract DE-FG02-03ER460097.  Use of the diffractometer at NSLS X1B was made possible by the Dutch NWO through its Spinozs program.

\bibliography{pcmo_sxrd_paper_condmatt}

\end{document}